%% file: main.tex
\documentclass[9pt,sigconf,letterpaper,screen]{acmart}
\def\BibTeX{{\rm B\kern-.05em{\sc i\kern-.025em b}\kern-.08emT\kern-.1667em\lower.7ex\hbox{E}\kern-.125emX}}
    
\usepackage{nicefrac}
\usepackage{siunitx}
\usepackage{array,framed}
\usepackage{booktabs}
\usepackage{
  color,
  float,
  epsfig,
  wrapfig,
  graphics,
  graphicx,
  subcaption
}
\usepackage{textcomp}
\usepackage{setspace}
\usepackage{latexsym,fancyhdr,url}
\usepackage{enumerate}
\usepackage{algorithm2e}
\usepackage{algpseudocode}
\usepackage{graphics}
\usepackage{xparse} % argument parsing -- \edist
\usepackage{xspace}
\usepackage{multirow}
\usepackage{csvsimple}
\usepackage{balance}
\usepackage{algpseudocode}
\usepackage{
  tikz,
  pgfplots,
  pgfplotstable
}
\usepackage{hyperref}

\usetikzlibrary{
  shapes.geometric,
  arrows,
  external,
  pgfplots.groupplots,
  matrix
}

\pgfplotsset{compat=1.9}
\usepackage{mathtools,}

\DeclareMathAlphabet{\mathcal}{OMS}{cmsy}{m}{n}
\DeclareGraphicsExtensions{%
    .png,.PNG,%
    .pdf,.PDF,%
    .jpg,.mps,.jpeg,.jbig2,.jb2,.JPG,.JPEG,.JBIG2,.JB2}

\input{defs}
\begin{document}
%\fontfamily{lmr}\selectfont
% \def\thetitle{A Practical Way to Generate Strong Keys from Noisy Data}
\fancyhead{}
\def\thetitle{A Blockchain-based two Factor Honeytoken Authentication System}
\title{\thetitle}

\author{Vasilis Papaspirou}
\affiliation{\small{University of West Attica}}
\author{Leandros Maglaras}
\affiliation{\small{Edinburgh Napier University}}
\author{Ioanna Kantzavelou}
\affiliation{\small{University of West Attica}}
\author{Naghmeh Moradpoor}
\affiliation{\small{Edinburgh Napier University}}
\author{Sokratis Katsikas}
\affiliation{\small{NORCICS}}

\date{}

\input{abstract}
\maketitle

\vspace{-12pt}
\section{Introduction}

Since digital transformation revises all activities in enterprises and embeds them into the digital world, advanced security solutions, including novel authentication mechanisms, are needed to address emerging threats. Two-factor Authentication (2FA)  methods have been introduced to enhance the security of traditional authentication structures but face significant security issues due to the complexity of the generated systems \cite{papathanasaki2022modern}. 

An innovative honeytoken solution approach, proposed in \cite{papaspirou2021novel}, overcomes many of the drawbacks associated with traditional methods and enhances their robustness. However, the problem of immutability exposes these methods to tampering attacks. Therefore, there is a need to address tampering issues and secure the authentication procedure by incorporating innovative technologies \cite{edwards2022ffda}.

Blockchain technology is a decentralized, distributed ledger that securely links together hashes of blocks containing information. It offers a potential solution to certain security problems by ensuring immutability, anonymity, consistency, and control \cite{ferrag2019blockchain}. Any authentication mechanism requires the transfer of sensitive information, which is vulnerable to attacks. Blockchain can act as an intermediary between two parties participating in an authentication procedure. In this section, we provide a brief overview of some recent state-of-the-art solutions that combine 2FA mechanisms with blockchain technology.

%Blockchain-based authentication methods have gained recently attention since they combine decentralized solutions, achieve mutual trust, and help avoid additional overhead. Moreover, blockchain can provide authentication logs for auditing purposes, verify One Time Passwords (OTP) that are produced by the authentication mechanisms, and ensures the integrity of data stored in it. 

The authors in \cite{park2018totp} proposed a 2FA method that incorporates Hyperledger blockchain technology and is based on time-based one-time passwords (TOTPs). The underlying idea of this model is to authenticate the user in two stages. The first stage involves the client and the server, while the second stage includes a peer that verifies the access token and the transaction between the client and the server. However, this solution is vulnerable to token collision and Man-in-the-Middle (MITM) attacks, and it cannot guarantee the security requirements of critical applications, such as web banking systems.

Several authors have proposed the inclusion of biometric characteristics and storing some or all of the information on public or private blockchains \cite{hammudoglu2017portable, zhou2018biometric, lu2018privacy}. The authentication of users can occur using this stored information along with traditional usernames and passwords. However, these methods are vulnerable to server compromise attacks, tampering, and impersonation attacks. In another recent work \cite{bandara2022casper}, the storage of users in mobile identity wallets was proposed to capture and verify user identity proofs. However, this solution has not yet incorporated a second factor of authentication.

More recently, the authors in \cite{catalfamo2021microservices} identified the weaknesses of a typical 2FA  system that is vulnerable to denial of service (DOS) and MITM attacks. They proposed a blockchain-based OTP (One-Time Password) protocol. This work builds upon other recent solutions \cite{ buccafurri2020securing, alharbi2019two}, where the blockchain is utilized to verify the user and generate the OTP. The proposed protocol proves to be user-friendly and decouples the direct interaction of each user with the blockchain.

All of the aforementioned works integrate a 2FA system with a blockchain for verification and authentication purposes. However, they still remain vulnerable to various threats that are typical of 2FA systems. In contrast, our recent two-factor honeytoken authentication solution \cite{papaspirou2022security} has demonstrated robustness against many common attacks. By incorporating blockchain technology, our solution is further strengthened, particularly against tampering attacks. Furthermore, implementing this honeytoken method via Blockchain ensures the method's immutability.
\vspace{-12pt}
\section{The B2FHA Platform}

The main drawback of a traditional centralized OTP generating method is that it consolidates all the procedures into a single service, thereby leaving applications vulnerable to MITM attacks. In such an incident, an attacker could potentially infiltrate the system and easily impersonate any user's login, generating OTPs as well.

The proposed Blockchain-based Two-Factor Honeytoken Authentication (B2FHA) system incorporates Blockchain technology into the recently introduced honeytoken architecture \cite{papaspirou2021novel}. In this system, the user selects one OTP from the $N$ available options as the valid one. For experimental purposes, the prototype currently supports N equals to 3, but it can be extended to include ten or more OTPs per user per session. The B2FHA integrates Blockchain technology to ensure immutability, simplicity, and efficiency compared to a typical centralized 2FA mechanism. The B2FHA system is characterized by three highlighted features: \textit{Honeytokens}, \textit{Blockchain}, and a  \textit{new validation process} that incorporates the proposed method.

%The features of the B2FHA  are highlighted below:
%\begin{itemize}
%  \item Honeytokens: Every user has a honeytoken used in the validation process, which is based on honeywords method.
%  \item Blockchain: It enhances the security of data when transferred.
%  \item Validation process: It validates the user honeytoken to successfully authenticate him/her.
%\end{itemize}

The validation process requires each user's honeytoken, while the Blockchain secures the connection and protects the transmission of sensitive data. An additional layer of security is provided through honeytoken encryption. During the registration phase, the user specifies the honeytoken to be used during the login phase. The application generates $N$ OTPs, and the specified honeytoken is stored for the corresponding user. When the OTP is requested for user authentication, the honeytoken is transferred via the Blockchain to the validation process, where the authentication mechanism determines the user's authenticity.

\begin{figure}[h]
\centering{\includegraphics[scale=0.3]{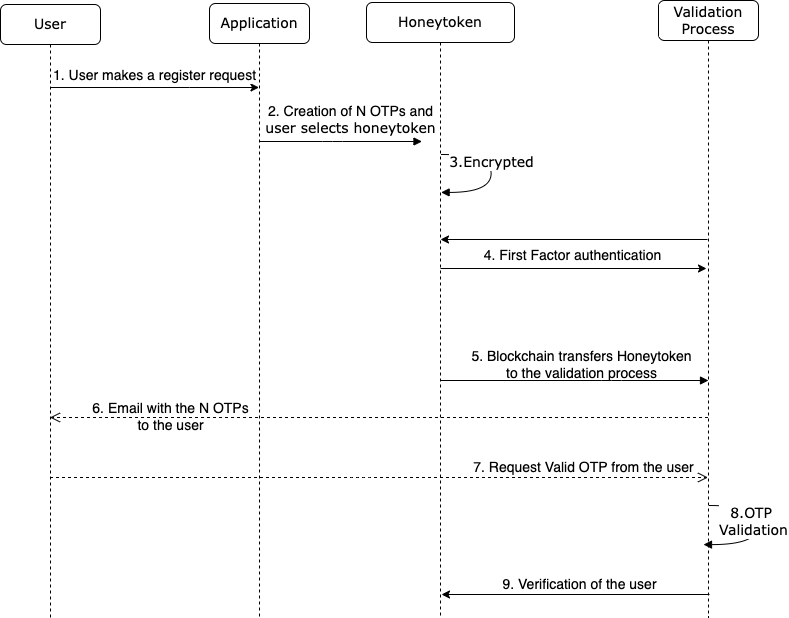}}
\caption{B2FHA system sequence diagram}
\label{fig:Fig1}
\end{figure}

Figure \ref {fig:Fig1} illustrates the steps of the B2FHA method in a sequence diagram. The diagram portrays processes, objects, and the messages exchanged between them, depicting the lifecycle of the B2FHA method from honeytoken generation to the completion of the validation process.

\vspace{-8pt}

\begin{enumerate}
    \item The user sends a register request to the application.
    \item The application creates $N$ OTPs, and the user specifies which honeytoken will be the valid one.
    \item The specified honeytoken is encrypted and stored.
    \item First-factor authentication request (username, password).
    \item Private Blockchain securely transfers the encrypted honeytoken to the validation process.
    \item An email with the $N$ OTPs is sent to the user.
    \item Valid OTP request to be sent to the validation process.
    \item The validation process checks the OTP entered by the user and also checks the honeytoken.
    \item The validation verifies weather a valid OTP was entered. A fake OTP locks the account.
\end{enumerate}

Unlike centralized 2FA methods, B2FHA divides all phases into separate components. The B2FHA method receives the user's honeytoken and securely delivers it through the Blockchain to the validation process. No OTPs are stored, and no sensitive user information is exposed during the OTP request. The pseudocode for the validation process is provided in Algorithm \ref{alg:cap}.

\begin{algorithm}
\caption{Validation process}\label{alg:cap}
user(username) and user(password)$\gets Login$\;

\eIf{$user\gets True$}
{
    $user\gets OTP request$\;
    }{
    $user\gets redirect Login$\;
    
    \eIf{$user(token)\gets Valid OTP$}
    {
    $user\gets Authenticated$\; 
    }{
    \eIf{$user(token)\gets fake OTP$}
    {
    $user\gets Locked$\;}
    {
    
    \If{$user(token)\gets mistyped OTP$}
    {
    $user\gets redirect Login$\;}}
    
    }
    }

\end{algorithm}

\vspace{-8pt}
\section{Evaluation}

Compared to a conventional 2FA method, the incorporation of blockchain into the honeytoken-based method serves as an additional component of the B2FHA system. Consequently, initial tests primarily focused on evaluating the overhead generated by the private Blockchain versus the implementation without Blockchain.

We examined and compared the times taken for honeytoken transfer to the validation endpoint in both systems. The tests were conducted on a local server with the following specifications: Ryzen 5 2600 3.4 GHz, 6-core CPU, 16 GB RAM, running on Windows 10. For both systems being tested, the tests were performed by setting a breakpoint at the beginning of OTP generation and stopping when the honeytoken reached the validation process.

The application of blockchain technology resulted in significantly faster processing times compared to the non-blockchain application, contrary to our initial expectations. As shown in Table \ref{long}, the blockchain-enabled solution has enhanced honeytoken verification efficiency and reduced the overall processing time.

This result underscores the potential of blockchain technology in streamlining and accelerating the verification procedure within the Honeytoken system. While the non-blockchain application offered advantages such as simplicity and a seamless user experience, the inclusion of blockchain outperformed in terms of processing speed. These findings suggest that integrating blockchain into the Honeytoken system could enhance performance and operational efficiency.

\vspace{-8pt}

\section{Conclusions}

Traditional 2FA methods often fall short in protecting systems from various attacks, including tampering attacks. However, the integration of blockchain technology into a 2FA honeytokens-based method introduces a novel mechanism. Through testing the B2FHA system, the observed results demonstrate that blockchain can strengthen the authentication mechanism without introducing additional overhead.

\begin{center}
\begin{table}[h!]
\caption{Time comparison when sending honeytoken to validation process\label{long}}

\begin{tabular}{ | m{8em} | m{1.3cm}| m{1.3cm} | m{1.3cm} | } 
  \hline
  Application & First & Second & Third  \\ 
  \hline
  Without Blockchain  & 0.3738 ms & 0.3492 ms & 0.3513 ms \\ 
  \hline
  With Blockchain & 0.00016 & 0.00017 & 0.00013\\ 
  \hline
\end{tabular}
\label{table:1}
\end{table}

\end{center}


\bibliographystyle{ACM-Reference-Format}
\bibliography{bib}

% % --- Appendix ---%

\end{document}

%% file: defs.tex
\usepackage{xparse}
\newcommand{\bnm}{\begin{newmath}}
\newcommand{\enm}{\end{newmath}}

\newcommand{\bea}{\begin{eqnarray*}}%
\newcommand{\eea}{\end{eqnarray*}}%

\newcommand{\bne}{\begin{newequation}}
\newcommand{\ene}{\end{newequation}}

\newcommand{\bal}{\begin{newalign}}
\newcommand{\eal}{\end{newalign}}

\newenvironment{newalign}{\begin{align}%
\setlength{\abovedisplayskip}{4pt}%
\setlength{\belowdisplayskip}{4pt}%
\setlength{\abovedisplayshortskip}{6pt}%
\setlength{\belowdisplayshortskip}{6pt} }{\end{align}}

\newenvironment{newmath}{\begin{displaymath}%
\setlength{\abovedisplayskip}{4pt}%
\setlength{\belowdisplayskip}{4pt}%
\setlength{\abovedisplayshortskip}{6pt}%
\setlength{\belowdisplayshortskip}{6pt} }{\end{displaymath}}

\newenvironment{newequation}{\begin{equation}%
\setlength{\abovedisplayskip}{4pt}%
\setlength{\belowdisplayskip}{4pt}%
\setlength{\abovedisplayshortskip}{6pt}%
\setlength{\belowdisplayshortskip}{6pt} }{\end{equation}}

\newcounter{ctr}

\newcounter{mytable}
\def\mytable{\begin{centering}\refstepcounter{mytable}}
\def\endmytable{\end{centering}}

\newcounter{myfig}
\def\myfig{\begin{centering}\refstepcounter{myfig}}
\def\endmyfig{\end{centering}}

\newlength{\saveparindent}
\newlength{\saveparskip}
\newcommand{\E}{{\rm I\kern-.3em E}}

\renewcommand{\eqref}[1]{\mbox{Equation~(\ref{#1})}}

\def \part {part}

\renewcommand{\paragraph}[1]{\vspace*{6pt}\noindent\textbf{#1}\;}

\def \blackslug{\hbox{\hskip 1pt \vrule width 4pt height 8pt
    depth 1.5pt \hskip 1pt}}
\def \qed{\quad\blackslug\lower 8.5pt\null\par}
\newcounter{mynote}[section]

\newcommand\ignore[1]{}

\newcounter{rcnote}[section]

\newcounter{mrnote}[section]

\newcounter{fknote}[section]

\newcounter{anote}[section]

\DeclareMathSymbol{\mlq}{\mathord}{operators}{``}
\DeclareMathSymbol{\mrq}{\mathord}{operators}{`'}

\newcommand{\rhf}[2]{R_{f, \gamma}}

\DeclareDocumentCommand{\edist}{o o}{
  \ensuremath{
    \IfNoValueTF{#1}{{d}}{{\sf d}(#1,#2)}
  }
}

\newcommand{\olrk}[1]{\ifx\nursymbol#1\else\!\!\mskip4.5mu plus 0.5mu\left(\mskip0.5mu plus0.5mu #1\mskip1.5mu plus0.5mu \right)\fi}

\NewDocumentCommand{\indseq}{ O{1} O{r} }{{#1}\ldots {#2}}

%% file: abstract.tex
\begin{abstract}
 This paper extends and advances our recently introduced two-factor Honeytoken authentication method by incorporating blockchain technology. This novel approach strengthens the authentication method, preventing various attacks, including tampering attacks. Evaluation results demonstrate that integrating blockchain into the Honeytoken method can enhance performance and efficiency.
\end{abstract}